# DEPENDENCE OF THE ENERGY OF MOLECULES ON INTERATOMIC DISTANCE AT LARGE DISTANCES

## I.A. STEPANOV

*Juglas 1a -20, Riga, LV-1024, Latvia*

Earlier it was supposed that the energy of molecules increases monotonously with interatomic distance at large distances. However, dissociation of molecules (for example, $Te_2 \rightarrow 2Te$) often is a chemical reaction. According to chemical kinetics, chemical reactions overcome a potential barrier. Therefore, there must be a barrier at the energy – distance curve. Earlier it has been supposed that quantum chemical methods give a wrong result at big distances if the wave function does not turn to zero. It is shown that it must not obligatory turn to zero. The wave function can be a piecewise function.

## 1. Introduction

According to the traditional point of view, the energy of molecules depends on interatomic distance according to Curve 1, Fig. 1 (the energy of independent atoms is supposed to be zero). This dependence has the following disadvantages. Dissociation of molecules (for example, $F_2 \rightarrow 2F$) often is a chemical reaction. According to chemical kinetics, chemical reactions overcome a potential barrier. This barrier is absent at Curve 1.

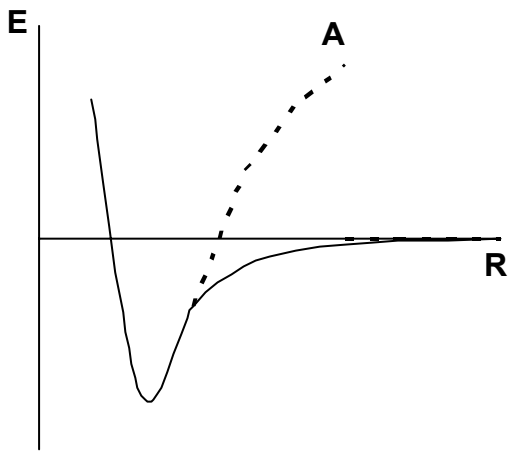

*Figure 1.* Dependence of the energy of molecule on the distance between atoms: 1 – traditional theory, 2 – according to this paper. Point A is the point where the bond begins to fail.

It is a very strong argument against Curve 1. Rupture is a transition from the less stable state to the more stable state, from the state with higher energy to the state with lower energy. According to non-equilibrium thermodynamics, the system being deflected greatly from equilibrium looses steadiness and the system changes to a qualitatively new steady state (with lower energy) [1]. In [2-4] it has been shown that failure of molecules during stretching of solids happens like this: at strong stretching of interatomic bonds, molecules loose stability and turn to a qualitatively new steady state with lower energy, resp. ruptured molecules. One can assume that dissociation of a single molecule also happens like this. According to the Curve 1 (see Figure 1), the energy must not be released during separation of interatomic bond. However, it has been found experimentally that during stretching of solids, chemical bond ruptures lead to micro-heating of substance to a few hundred grades [5].

## 2. Theory

The dependence of the energy of diatomic molecule on the distance between the atoms must be the following one: the Curve 2 (Figure 1). During stretching of molecule, its energy becomes larger than that of independent atoms and in the point A the molecule becomes greatly unstable and turns to the dissociated state. There are two possibilities: from the point A the curve tends smoothly to zero, or transition to the dissociated state happens by a jump. With this the energy of elastic stretching is released.

Pay attention that the function E(R) defined at $R_0 \leq R < R^*$ ($E(R_0)$ is minimal, $E(R^*) = 0$, Curve 2, Fig. 1) at $R \to \infty$ does not turn to zero. The energy E(R) at $R \to \infty$ has the sense of the energy of fictitious molecule being stretched to the infinitive distance. Dependence E(R) near the bottom of the potential well is found experimentally, behaviour of E(R) at big distances is an invention of physicists. It has been supposed that E(R) at $R \to \infty$ turns to zero. It is not obvious. Molecule must not obligatory fail if E(R) = 0.

In [6] the H2+ ion has been solved exactly taking into account that the energies of electron - nuclei interaction are

$$E_1 = 1/2 K r_1^2, \tag{1}$$
$$E_2 = 1/2 K r_2^2 \tag{2}$$

where K is the coefficient of proportionality, $r_i$ is the distance between nucleus and the electron, and the energy of nuclei interaction is

$$E_3 = \lambda/R^2, \lambda > 0, \tag{3}$$

where R is the inter-nuclear distance. According to this calculation, the energy of the ground state is

$$E(R) = 3/2 \cdot (2K/m)^{1/2} + KR^2/4 + \lambda/R^2, \tag{4}$$

where m is the hydrogen atom nucleus mass. The attraction force in such ion is greater than that in the real one, and the repulsion force is less. It means that the binding energy of such ion is bigger and dE/dR for it is bigger than that for the real ion. Let us build the following H2+ ion model: near the bottom of the potential well energy is described by (4), at bigger distances it is described by the same equation but K and λ depends on R. In this model E(R) behaves like the Curve 1, Figure 1. By fitting of K(R) and λ(R) in (4) one can ascertain that the depth of the potential well is the same as that with the real H2+. In such model dependence E(R) is stiffer and must reach zero by smaller R than that of the real ion. It is a contradiction: the ion with larger binding energy fails earlier than that with the smaller one. Therefore, the initial supposition that bond failure happens at E = 0, is not true. The rupture of chemical bond begins in the real ion at $R_A$, $E(R_A) > 0$, but the bond rupture in the model ion begins at $R_M > R_A$, $E(R_M) > E(R_A)$.

## 3. Results and Discussion

Earlier it was supposed that quantum chemical methods give a wrong result at big distances if the wave function does not turn to zero at $R \to \infty$. It is necessary to make the conclusion that the wave function must not turn to zero. This result explains the paradox: experimental dissociation energies usually are much bigger than theoretical ones [7, 8]. Mulliken in [9] predicted the existence of maxima at the energy – distance curves: „Here a theoretical calculation by Pauling on $He_2^{++}$ should also be mentioned, in which he found that the potential energy curve should have a pronounced maximum … In view of these several examples where recognition of the existence of maxima has been practically *forced* on us, it seems likely that such cases may prove relatively frequent if we try to *seek* them. After all, maxima in potential energy curves are obvious necessities for polyatomic molecules, in view of the existence of activation energies, so that their occurrence also for diatomic molecules is not at all shocking."

In astrophysics the barriers at Curve 2 (Figure 1) are found experimentally for some molecules. They are barriers of chemical reaction [10]. Many dependencies in Nature are piecewise functions. For example, the dependence of strength of solids is a piecewise function of length and diameter of the specimen, scatter in strength in solids is a piecewise function of strength or the time the load is withstood [2–5], the chemical reaction rate constant and equilibrium constant are piecewise functions of temperature [11–12]. It is very possible that the wave function is also a piecewise function of its arguments:

$\psi(R) = \psi_1(R), \quad R_0 \leq R < R_1;$
$\psi(R) = \psi_2(R), \quad R_1 \leq R < R_2;$
$\psi(R) = \psi_3(R), \quad R_2 \leq R < R_3;$
………………………….. . (5)

One sees that $\psi_1(R)$ must not turn to 0 when R turns to infinity, even if there is no hump at the Curve 1 (Figure 1). Earlier, the authors added additional terms to $\psi_1(R)$ to secure $\psi_1(R) \to 0$ at $R \to \infty$. This trick seems doubtful now.